\documentclass[sigconf]{aamas}  

\usepackage{booktabs}
\usepackage[ruled,vlined,linesnumbered, resetcount]{algorithm2e}
\usepackage{makecell}
\usepackage{subcaption}
\usepackage{enumitem}

\AtBeginDocument{%
  \providecommand\BibTeX{{%
    \normalfont B\kern-0.5em{\scshape i\kern-0.25em b}\kern-0.8em\TeX}}}
    
\usepackage{booktabs}    
\usepackage{flushend} 

\setcopyright{ifaamas}  
\copyrightyear{2020} 
\acmYear{2020} 
\acmDOI{} 
\acmPrice{} 
\acmISBN{} 
\acmConference[AAMAS'20]{Proc.\@ of the 19th International Conference on Autonomous Agents and Multiagent Systems (AAMAS 2020)}{May 9--13, 2020}{Auckland, New Zealand}{B.~An, N.~Yorke-Smith, A.~El~Fallah~Seghrouchni, G.~Sukthankar (eds.)}  

\begin{document}
\title[Influence maximization in unknown social networks]{Influence maximization in unknown social networks:
Learning Policies for Effective Graph Sampling}

\author{Harshavardhan Kamarthi$^{1}$ Priyesh Vijayan$^{2}$ Bryan Wilder$^{3}$ Balaraman Ravindran$^{1,4}$ Milind Tambe$^{3}$}
\affiliation{$^1$Dept. of Computer Science, Indian Institute of Technology Madras\\
$^2$School of Computer Science,McGill University and Mila\\
$^3$Center for Research on Computation and Society, Harvard University\\
$^4$Robert Bosch Centre for Data Science and AI, Indian Institute of Technology Madras
}

\begin{abstract}
A serious challenge when finding influential actors in real-world social networks, to enable efficient community-wide interventions, is the lack of knowledge about the structure of the underlying network. Current state-of-the-art methods rely on hand-crafted sampling algorithms; these methods sample nodes and their neighbours in a carefully constructed order and choose opinion leaders from this discovered network to maximize influence spread in the (unknown) complete network. 

In this work, we propose a reinforcement learning framework to discover effective network sampling heuristics by leveraging automatically learnt node and graph representations that encode important structural properties of the network.
At training time, the method identifies portions of the network such that the nodes selected from this sampled subgraph can effectively influence nodes in the complete network. The output of this training is a transferable, adaptive policy that identifies an effective sequence of nodes to query on unseen graphs. The success of this policy is underpinned by a set of careful choices for embedding local and global information about the graph, and providing appropriate reward signals during training. We experiment with real-world social networks from four different domains and show that the policies learned by our RL agent provide a 7-23\% improvement over the current state-of-the-art method.
\end{abstract}

\maketitle

\setlength\abovecaptionskip{3.6pt}

\section{Introduction}

Social network interventions are used across a wide variety of domains to disseminate information or inspire changes in behavior; application areas range from substance abuse \cite{valente2007identifying}, to microfinance adoption \cite{banerjee2013diffusion}, to HIV prevention \cite{yadav2018bridging,wilder2018end}. Such processes are computationally modelled via the \emph{influence maximization} problem, where the goal is to select a subset of nodes from the network to spread a message, such that the number of people it eventually reaches is maximized. Several algorithmic approaches have been proposed for influence maximization \cite{kempe2005influential,jung2012irie,chen2010scalable,li2018influence}.

However, real-world applications of influence maximization are often limited by the high cost of collecting network data. In many domains, for instance, those arising in public health, a successful intervention requires information about the face-to-face interactions with the members of a population. This information is typically gathered via in-person surveys; conducting such surveys requires substantial effort on the part of the organization deploying an intervention. We are motivated in particular by the problem of using influence maximization for HIV prevention among homeless youth, where algorithms have been successfully piloted in real-world settings \cite{yadav2018bridging,wilder2018end}. In the HIV prevention domain, gathering the social network of the youth who frequent a given homeless centre requires a week or more of effort by social workers, which is not feasible for a typical community agency. 

The method proposed in \cite{wilder2018end} addresses this by introducing the CHANGE algorithm which uses a simple yet effective sampling method: surveying random nodes and one of their neighbours to discover parts of the social network. Each node that is surveyed reveals its neighbours, allowing us to observe a subgraph of the entire social network. The subgraph is used to pick influential set of nodes by applying a influence maximization algorithm on it. Thus the discovered subgraph acts as a surrogate to the entire network for deciding on influential actors. Existing network discovery algorithms are entirely hand-designed \cite{wilder2018maximizing,wilder2018end}, typically aiming to exploit a specific property of graphs such as community structure or the friendship paradox \cite{feld1991your} to get the surrogate graph.

This raises the question of whether it is possible to \textit{automatically} train an agent for network discovery; exploring from the entire space of network discovery policies may yield more effective results than hand-engineered approaches. Hence, we employ reinforcement learning (RL) to discover better heuristics on a given network dataset. Instead of hard-coding a reliance on particular structural network features (as in previous works \cite{wilder2018end, wilder2018maximizing}), our agent learns from scratch how to represent the social network in a way that is conducive to effective sampling policies. This allows it to exploit additional structural properties that were not used in previous approaches, ultimately enabling better performance.  

In particular, our approach leverages the availability of historical network data from similar populations (for instance, when deciding which youths to survey for an HIV prevention intervention, we can deploy policies learnt from training using networks gathered at other centres).As a consequence, our agent learns more nuanced policies which can both be fine-tuned better to a particular distribution of graphs and which can adapt more precisely over the course of the surveying process itself (since the trained policy is a function of a graph discovered so far). Even if training networks are not available from the specific domain of interest, we show that comparable performance can also be obtained by training on standard social network datasets or datasets generated synthetically using community structure information and transferring the learned policy unchanged to the target setting. 


\noindent The main contributions of our work can be summarized as follows:
\begin{enumerate}[leftmargin=*]
    \item We formulate the process of network discovery for influence maximization as a sequential decision problem via a Markov Decision Process (MDP). To solve this MDP, we propose a neural network architecture called Geometric-DQN and a training algorithm that uses Deep Q-learning to learn policies for network discovery by extracting relevant graph properties from the training dataset. To the best of our knowledge, this is the first work to use deep reinforcement learning for network discovery to aid influence propagation in unknown networks.   
    
    \item One key feature of Geometric-DQN is its ability to learn both global state representations for the entire discovered graph and local action representations for the individual nodes; capturing information at both scales allows our agent to learn nuanced policies. Our RL algorithm can accommodate an arbitrary number of actions (nodes) that can vary depending on the state (graph). We also enable training across multiple graphs by designing an appropriately scaled reward scheme that is applicable across different kinds of networks. This allows Geometric-DQN to benefit from training on multiple similar graphs resulting in superior performance compared to the baselines.


    \item Motivated by the benefits of training on multiple networks, we propose to use synthetically generated graphs from a similar population to augment existing networks. This allows the model to obtain improved performances, especially when there are limited number of training networks. To this end, we design algorithms that leverage community structure from available training graphs to generate graphs with similar properties.
    
    \item We experimentally evaluate our RL based network discovery algorithm on social networks from four different domains. Our model outperforms previous state-of-the-art sampling approaches by a substantial margin of 7-23\%.
\end{enumerate}
\label{sec:inflmodel}
\section{Problem Description}
\label{sec:ps}
Our problem is ultimately motivated by goal of choosing a set of seed nodes in a social network who will disseminate information to as many other nodes as possible. To model information diffusion over the network, we use the standard independent cascade model (ICM) \cite{kempe2003maximizing}, which is the most commonly used model in the literature. In the ICM, every node is either active or inactive. At the start of the process, every node is inactive except for the seed nodes, $S$. The process unfolds over a series of discrete time steps. At every step, each newly activated node attempts to activate each of its inactive neighbors and succeeds with some probability $p$. The process ends when there are no newly activated nodes at the final step..

\begin{figure}[ht]
    \centering
    \includegraphics[width=0.4\textwidth]{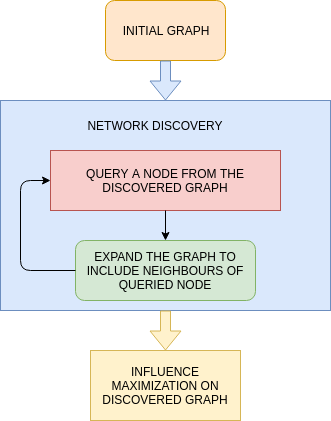}
    \caption{Network discovery for Influence maximization}
    \label{fig:pipe}
\end{figure}

Our algorithm is concerned with discovering a subgraph of the social network such that the $|S|$ seed nodes chosen from the discovered subgraph maximize the number of activated nodes at the end of the process. The subgraph is discovered as follows: at each step, we query a node from the already discovered subgraph. The queried node then reveals its neighbors, expanding the discovered subgraph. This process goes on for fixed number of steps after which we use the final discovered subgraph as input to an influence maximization algorithm. This process is illustrated in Figure \ref{fig:pipe}.

We can define the problem more formally as a sequential decision problem.
Let the entire unknown graph be $G^*=(V^*,E^*)$. Let $X\subseteq V^*$ denote a vertex set. Let $G[X]$ denote a sub-graph of $G^*$ induced by $X$. Let $V(G)$ be the vertex set of a graph $G$ and $E(G)$ be edge set of $G$. Let $N_G(u)$ be neighbors of vertex $u$ in a graph $G$. 
$E(X,Y)$ be all direct edges that connect a node in $X$ and a node in $Y$.

Initially, we are given a set of seed nodes, $S$, whose connections are observed. When no prior information is available, these seeds could be obtained by querying nodes at random. The agent has a budget of $T$ queries to gather additional information. When we query a node, we discover the neighbours of the queried node. Let $G_t$ be the sub-graph discovered after $t$ queries with  vertex set, $V_t = V(G_t)$.
Let $G_0=(S\cup N_{G^*}(S), E(S,N_{G^*}(S)))$. During the ($t+1$)\textsuperscript{th} query, we choose a node $u_{t}$ from $G_{t}$ and observe $G_{t+1}=(V_t \cup N_{G^*}(u_t), E(G_t)\cup E(N_{G^*}(u_t),\{u_t\}))$. For any observed graph, we can use the \textit{greedy influence maximization algorithm} \cite{kempe2005influential} (which assumes that the entire graph is known) as an oracle to determine the best set of nodes to activate based on the available information. Let $\mathcal{A}=\mathcal{O}(G)$ be the output of this oracle on graph $G$, and let $I_G(\mathcal{A})$ be the expected number of influenced nodes in $G$ on choosing $\mathcal{A}$ as the set of initial active nodes.
The task is to find a sequence of queries $(u_0,u_1,\dots,u_{T-1})$ such that the discovered graph $G_T$ maximizes $I_{G^*}(\mathcal{O}(G_T))$, i.e, we need to discover a subgraph $G_T$ such that the nodes selected by $\mathcal{O}$ in $G_T$ maximizes the number of nodes influenced in the entire graph, $G^*$.

\begin{figure*}[!ht]
    \centering
    \includegraphics[width=0.7\textwidth]{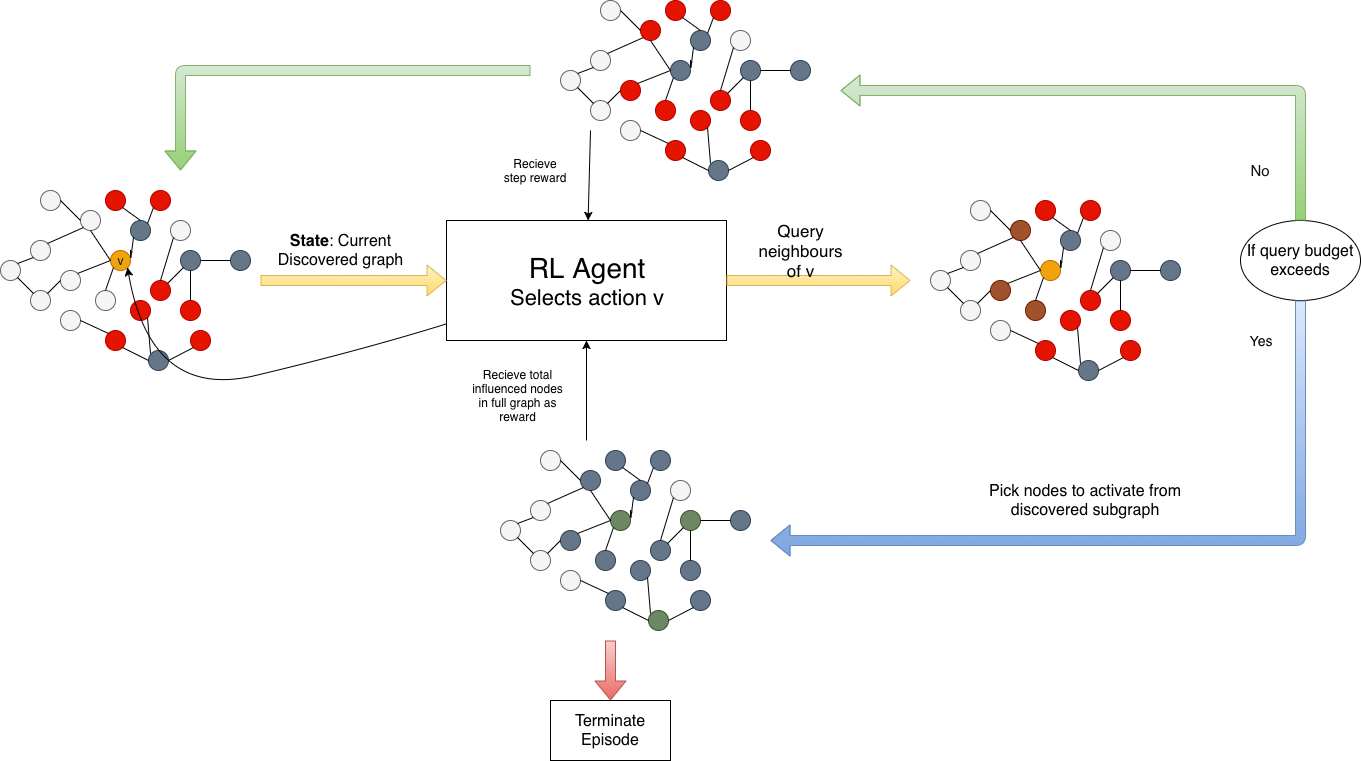}
    \caption{RL agent pipeline. (Red: Nodes in current action set, Yellow: Node selected by RL agent, Grey: Other nodes from discovered graph, Green: Nodes picked to activate by greedy influence maximization algorithm)}
    \label{fig:piperl}
\end{figure*}

There is no existing method to directly optimize $I_{G^*}(\mathcal{O}(G_T))$ given $G_0$. Therefore, we employ reinforcement learning by converting the problem into a Markov Decision Process (MDP), to explore different heuristics and generate improved sampling policies over time. Since we use deep reinforcement learning methods that require the states and actions to be real valued vectors, we need to extract vector representations that summarize important properties of individual nodes or of the graph as a whole. 
We show how we can can automatically learn effective representations in tandem with process of discovering policies via a Q-learning based algorithm.

\section{Methodology}

Previous work has used entirely hand-designed heuristics for network discovery. In order to automatically produce more effective agents for the problem, we introduce a series of methodological contributions. \textit{First}, we provide a MDP formulation of the problem, with appropriately defined rewards to enable training across multiple graphs. \textit{Second}, we propose a step-reward function to alleviate reward sparsity. \textit{Third}, we propose a combination of local and global embeddings as the state and action representation, which both solves unique difficulties in the way our problem's action space and provides more nuanced contextual information to our agent. Previous work on graph reinforcement learning \cite{Khalil2017LearningCO} did not use a distinct global representation, which we found to be an important part of success in our domain. \textit{Fourth}, we propose a data augmentation technique based on fitting a probabilistic model and generating additional synthetic graphs, allowing our agent to perform well when real-world network data is scarce. The following sections detail these contributions. 

\subsection{Markov Decision Process Formulation of Network Discovery problem}
\label{sec:mdpf}

We start by formalizing the network discovery problem as a MDP. 

\textbf{State:} The current state is the discovered graph $G_t$.

\textbf{Actions:} Given a sub-graph $G_t$, we can query any of the nodes in $G_t$ which are not yet queried. Thus, the action space is $V_t/\{S\cup_{i\leq t}u_i\} \forall t>0$ or is $N_{G^*}(S)$ if $t=0$.

\textbf{Rewards:} The reward we get after $T$ steps is the number of nodes influenced in the entire graph, $G^*$ using the discovered graph, $G_T$ which is $I_{G^*}(\mathcal{O}(G_T))$. We denote this reward, which the model receives at the end of an episode as $R_i$.
\paragraph{Rewards when using multiple graphs:}
When we train simultaneously on multiple graphs from the dataset, the task is no longer an MDP since the next state depends on which graph from the training dataset we are currently training on, which is unknown to the agent. Hence, the problem becomes a POMDP (Partially observable MDP) \cite{Lovejoy1991ASO} where the distribution over next state on choosing an action is also conditionally dependent on graph used along with previous state.

The range of rewards values are highly dependent on size and structure of the graphs considered. Therefore, we can't directly use this reward signal to train with multiple graphs simultaneously without appropriate reward normalization. 
We solve this problem by scaling the reward. Specifically, we compare the influence reward $I_{G^*}(\mathcal{O}(G_T))$ with reward from using the CHANGE algorithm \cite{wilder2018end}, a state-of-art algorithm that also serves as our baseline.
We also keep the range of reward between 0 and 1 to improve the stability of DQN training \cite{mnih2015human}. We normalize the influence reward as:
\begin{equation}
    R_s = \frac{I_{G^*}(\mathcal{O}(G_T)) - CHANGE(G^*)}{OPT(G^*)-CHANGE(G^*)}
    \label{eqn:scale}
\end{equation}
where $CHANGE(G^*)$ is the average number of influenced nodes when the graph is sampled using CHANGE \cite{wilder2018end} and $OPT$ is the number of influenced nodes when we select the active nodes given the knowledge of entire graph, i.e, $I_{G^*}(\mathcal{O}(G^*))$.

We use $CHANGE$ for bounding $R_s$ in Equation \ref{eqn:scale} because of three reasons. Firstly, $CHANGE$ is a powerful method for graph discovery and dominates other state-of-art sampling methods in terms of performance. Secondly, it is computationally inexpensive to simulate sampling according to $CHANGE$ on large networks. Hence we can pre-compute performance of $CHANGE$ for multiple training graphs before training the RL model. Finally, it is hard to find a feasible alternative to Equation \ref{eqn:scale} that uses network graph properties like size and density to estimate the range of influence rewards to scale it appropriately for different domains of networks.

\textit{Step-rewards:} While these rewards are well-scaled and translate across networks, they are only available at the end of each episode. Hence, we also add \emph{step-rewards} at each step to help alleviate reward sparsity and encourage the agent to learn to find larger graphs. The step-reward $R_{p,t}$ at time $t$ is given as:
\begin{equation}
    R_{p,t} = \frac{|V(G_t)| - |V(G_{t-1})|}{|V(G^*)|}
    \label{eqn:step}
\end{equation}
It was observed that while the step-rewards are about two orders of magnitude less than the influence reward in general, they are usually essential for initial stable learning. While we found that our method tends to discover larger graphs than the baselines, there was no correlation between a  model's performance and the size of the graph it discovers. We suspect that the step-rewards are important during initial phase of training, after which the final reward signal dominates.


\subsection{Representation learning for Geometric-DQN}
\label{sec:rep}

The MDP formulation presents us with challenges atypical of most reinforcement learning problems.
A social network is a very structured object that can vary in size and complexity. 
To condition the decision making process on the graph, we need a good vector representation of the graph.
The actions, which are nodes of the networks yet to be queried, need to be represented as vectors that encode structural information of the node in the context of the discovered network. \\ 

\noindent \textbf{Structure of Geometric-DQN network: }
The network discovery problem imposes new challenges on how representations are structured, which are not encountered in other RL problems. First, reinforcement learning problems usually have a fixed set of actions. Even when the action space is continuous, the range of action is known a priori and bounded. However, in this case, the size of action set (number of queryable nodes) depends on current graph. Second, the node representation is a function of the current state, i.e, the same node could have very different neighborhood properties and hence different embeddings in context of two different graphs. Therefore, the original DQN architecture \cite{mnih2015human}, which takes only the state representation as input and outputs action values, can't be used. To resolve these challenges, we input both state (high-level graph representation) and action representation (node embeddings) to the DQN and train it to predict the state-action value. This allows Geometric-DQN to augment the global information from graph embeddings with the neighbourhood information of node embeddings, adapting its actions to the information obtained so far. \\
\begin{figure}[!ht]
    \centering
    \includegraphics[width=.5\textwidth]{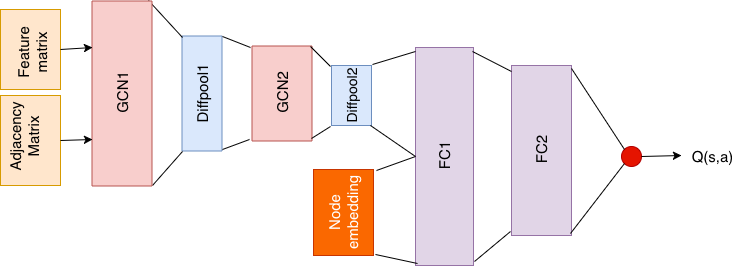}
    \caption{Geometric-DQN Architecture}
    \label{fig:arch}
\end{figure}

\noindent \textbf{Graph and Node Representation: }
Next, we discuss how we extract the state and action embeddings from structure of the network.

\textit{Background: Graph Convolutional Networks (GCNs)}: GCNs learn refined node features of a graph by passing and aggregating node features from the neighbors of each nodes. Given a graph $G$ with adjacency matrix $A \in \mathbb{R}^{n\times n}$ and a node feature matrix, $F^{(k-1)} \in  \mathbb{R}^{n\times d}$ in layer $k-1$ where $n$ is the number of nodes and $d$ is the number of features, a Graph Convolutional layer derives node features using a transformation function $F^{(k)}=M(A,F^{(k-1)};W^{(k)})$ where $W^{(k)}$ are weights of kth layer. 
In this work, we use the formulation in \cite{kipf2016semi} for GCN layers. Let $\Tilde{A}=A+I$ and $D=\sum_j\Tilde{A_{ij}}$.
\begin{equation}
    F^{(k)} = ReLU(D^{-\frac{1}{2}}\Tilde{A}D^{-\frac{1}{2}}F^{(k-1)}W^{(k)})
\end{equation}


\textit{Global state representation}: While GCNs refine node features by message passing and aggregating, we use differential pooling (Diffpool) \cite{ying2018hierarchical} to learn a global representation of the graph by aggregating node features in a hierarchical manner.
Diffpool learns hierarchical representations of the graph in an end-end differentiable manner by iteratively coarsening the graph, using GCNs as a building block.  Diffpool can be used to map a graph to a single finite dimensional representation by iteratively coarsening the input graph to a graph with a single node and extracting features for this new one node graph. 

We used DeepWalk embeddings \cite{perozzi2014deepwalk} as node features $\phi = F^{(0)}$ for the input layer. DeepWalk is a random-walk based node embedding model. It learns node representations that are similar to other nodes that lie within a fixed proximity on multiple random walks.

\textit{Local action representation}: We also utilized DeepWalk node embeddings, $\phi$ for representing nodes as action input in Q-network. We used deepwalk embeddings rather than randomly initializing $F^{(0)}$ since it was found to be vital for stable training of Geometric-DQN. Using GCN node features for actions resulted in unstable training as the input action space was non-stationary.
We used neighbourhood-based DeepWalk embeddings as it performed much better and was computationally less expensive than structure based encodings \cite{Epasto2019IsAS,Donnat2018LearningSN}.


\begin{algorithm}[!ht]

\SetAlgoLined
\SetKwInOut{Input}{Input}
\SetKwInOut{Output}{output}
\Input{Train Graphs $\mathcal{G}=\{G_1,G_2,\dots,G_K\}$, number of episodes $N$, Query budget $T$, number of random seeds $|S|$}

Initialize DQN $Q_{\theta}$ and target DQN $Q_{\theta'}$ with $\theta=\theta'$\;
Initialize Prioritized Replay Buffer $B$\;

\For{episode = 1 to $N$}{
    $G = sample(\mathcal{G})$\;
    $S = sample(G)$\;
    $V_0=S\cup N_{G}(S)$\;
    Initial graph is $G_0 = G[V_0]$\;
    $F_0^0 = DeepWalk(G_0)$ and $S_0=(F_0^{0},A_0)$\;
    $X \leftarrow N_G(S)$\;
    \For{t = 0 to $T-1$}{

        With probability $\epsilon$ select a random node $v_t$ from $X$ else select node $v_t \leftarrow \underset{v\in X}{argmax} Q_{\theta}(S_t,\phi(v))$\;
        
        Query node $v_t$ and observe new graph $G_{t+1}$\;
        
        Set $R\leftarrow R_{p,t}$ [Eqn. \ref{eqn:step}].\;
        
        \If{$t=T-1$}{
        $R\leftarrow I_{G^*}(\mathcal{O}(G_T))$\;
        
        $R_t = \frac{R - CHANGE(G^*)}{OPT(G^*)-CHANGE(G^*)}$
        }

        $F_{t+1}^0 = DeepWalk(G_{t+1})$ and $S_{t+1}=(F_{t+1}^{0},A_{t+1})$\;

        Add $(S_t,\phi(v_t),R_t,S_{t+1})$ to prioritized replay buffer \;
        
        Sample from $B$ and update $Q_{\theta}$\;
        
        $X$ is the set of nodes not yet queried in $G_{t+1}$\;

    }
    Update target network $Q_{\theta'}$ with parameters of $Q_{\theta}$ at regular intervals\;

}
\caption{Train Network}
\label{alg:train}
\end{algorithm}

\subsection{Model training and deployment}

For training, we can use single or multiple graphs. Algorithm \ref{alg:train} summarizes the training steps. 
At the start of every episode, we sample a graph at random from available training graphs (Line 4). Every time we expand the discovered graph, for timestep $t$, we compute node embeddings for all vertices in $G_t$ (Lines 8, 18). The Deepwalk representation for node $v$, denoted as $\phi(v)$, is a $d$ dimensional vector. Then we create the feature matrix $F_t^{0}\in \mathbb{R}^{|V_t|\times d}$ whose rows are the embeddings of nodes in $V_t$. $F_t^{0}$ is fed along with adjacency $A_t$ as input for Geometric-DQN. The state is represented by $S_t=(F_t^{0},A_t)$. For each of the nodes $v$ which are not queried yet until $t$, we get the state-value $Q(S_t,\phi(v_t))$ and choose the node $v_t$ that has maximum estimated state-value to query next (Line 12). We receive the reward as discussed and observe the new graph $G_{t+1}$ (Line 13).
We also store the experience $(S_t,\phi(v_t),R_t,S_{t+1})$ in replay buffer. Since we use prioritized replay \cite{schaul2015prioritized}, we also compute the TD error which is used to determine the importance of the experience in training.
Since the rewards are scaled (Equation \ref{eqn:scale}) we can combine experiences from different graphs, enabling the agent to generalize better.
After we train the DQN, we can deploy it on a new network after freezing the weights of Q-network. The deployment algorithm is similar to the training algorithm in that we compute Deepwalk embeddings for nodes at each step and find the next node to query based on the estimated state-values of all nodes which are not queried yet.

\subsection{Synthetic graph generation}
\label{sec:syn}
In many real domains, only a small number of actual networks are available. To alleviate this difficulty, we propose a data augmentation strategy which uses known structural properties of the family of networks to synthetically generate graphs for training. Even if a large number of training graphs are available, adding synthetic graphs provides the model with a richer training set and encourages robustness to small perturbations in the graph structure. 

Our strategy is based on Stochastic Block Models (SBMs), which originated in sociology \cite{holland1983stochastic} and can generate graphs that emulate structural properties seen in real world social networks.
SBMs generate random graphs containing densely connected communities. The individual communities generated are Erdos-Renyi graphs with edge probability $p_{in}$. The communities are interconnected sparsely by assigning an edge between each pair of nodes of different communities with a low probability $p_{out}$.

In this work, for each training graph, we construct a stochastic block model based on size of densely connected components of the graph called communities, determined using the louvain community detection algorithm \cite{blondel2008fast}.  The Louvain community detection algorithm maps each node of input graph to one of the communities. For each community, we find the maximum liklihood estimate $p_{in}$ since we know the actual number of edges within the community. Similarly, using edge information of actual graph, we find the maximum liklihood estimate $p_{out}$ for the allocated community structure. Thus, we have a SBM model with a distinct $p_{in}$ for each community.

For retweet networks, we make a small change to the SBM model, which we refer to as the Stochastic Star Model (SSM): We assume each community is a star-graph instead of a Erdos-Renyi graph (and estimate only $p_{out}$). This is justified by the observation that retweet networks usually \cite{sadikov2009information, leskovec2009community} looks like interconnected network of star graphs.
During training we generate  SBM or SSM graphs using parameters estimated on the real graphs.

\section{Experiment setup}
\noindent \textbf{Datasets:} \label{sec:data}
We evaluate the effectiveness of our model on datasets from four different domains: 1) Rural Networks - networks gathered by \cite{banerjee2013diffusion} to the study diffusion of micro-finance in Indian rural households, 2) Retweet Networks from twitter \cite{nr-aaai15}, 3) Animal Interaction Networks - networks are a part of the wildlife contact networks collected by \cite{davis2015spatial} to study the physical interactions between Voles, 4) Homeless Networks -from various HIV intervention campaigns organized for homeless youth in Los Angeles \cite{wilder2018end, wilder2018maximizing}. For each of the 4 families of networks, we randomly divide them into train and test data as shown in Table \ref{tab:split}. We use Algorithm \ref{alg:train} to train models on networks from training set and deploy them on the networks from test set for each network family.

\begin{table}[!ht]
\centering
\scalebox{0.88}{
\begin{tabular}{|l|c|c|}
\toprule
\textbf{Network category} & \textbf{Train networks} & \textbf{Test Networks}      \\
\midrule
\textbf{Rural}      & rural1,rural2           & rural3,rural4               \\
\textbf{Animal}     & voles1,voles2       &  voles3, voles4 \\
\textbf{Retweet}     & copen, occupy   & \makecell{isreal,damascus} \\
\textbf{Homeless} &a1,spy,mfp &  \makecell{b1,cg1,node4,\\mfp2,mfp3,spy2,spy3} \\
\bottomrule
\end{tabular}
}
\caption{Train and test split for different sets of networks}
\label{tab:split}
\end{table}

\begin{table*}[!htb]
\centering
\scalebox{0.9}{
\begin{tabular}{|l|cc|cc|cc|ccccccc|}
\toprule
& \multicolumn{2}{c|}{Rural} & \multicolumn{2}{c|}{Animals} & \multicolumn{2}{c|}{Retweet} & \multicolumn{7}{c|}{Homeless} \\
\midrule
Train\textbackslash{}Test & rural3      & rural4             & voles3        & voles4 & damascus          & israel & b1            & cg1           & node4         & mfp2          & mfp3          & spy2          & spy3\\\hline
\midrule
OPT         &  25.2 & 45.4 & 110.6 & 115.7 & 195.4 & 115.2 & 24.7 & 17.02 & 15.84 & 20.56 & 23.45 & 21.26 & 21.71\\
CHANGE             & 17.4 & 31.5 & 33.7 & 58.9 & 95.24 & 30.6 & 19.1 & 14.17 & 12.85 & 14.6 & 16.5 & 16.01 & 16.09\\
Geometric-DQN  & \textbf{18.95} & \textbf{35.7} & \textbf{45.8} & \textbf{80.2} & \textbf{119.3} & \textbf{43.6} & \textbf{20.2} & \textbf{14.98} & \textbf{13.9} & \textbf{15.95} & \textbf{17.7} & \textbf{17.4} & \textbf{18.2}\\

\bottomrule
\end{tabular}
}
\caption{Comparison of influence score of CHANGE and best performing Geometric-DQN model for each test network}
\label{tab:multi}
\end{table*}

\begin{table*}[!ht]
\centering
\scalebox{0.9}{
\begin{tabular}{|l|cc|cc|cc|ccccccc|}
\toprule
& \multicolumn{2}{c|}{Rural} & \multicolumn{2}{c|}{Animals} & \multicolumn{2}{c|}{Retweet} & \multicolumn{7}{c|}{Homeless} \\
\midrule
Train\textbackslash{}Test & rural3      & rural4             & voles3        & voles4 & damascus          & israel & b1            & cg1           & node4         & mfp2          & mfp3          & spy2          & spy3\\\hline
\midrule
Avg. individual         & \textbf{13.14} & 14.86                       & 13.07          & 29.05 & 11.58             & 8.16   & 4.76          & 12.51          & 25.75          & 11.74          & 7.18       & \textbf{21.40} & 31.61\\
Multiple             & 10.26        & \textbf{18.71}             & \textbf{14.84} & \textbf{35.21}  & \textbf{15.54}    & \textbf{15.37} & \textbf{14.29} & \textbf{22.11} & \textbf{35.12}  & \textbf{12.92} & \textbf{18.10}& 20.76          & \textbf{35.77}\\
\bottomrule
\end{tabular}
}
\caption{Comparing average \emph{improve percent} using individual network with that for training with multiple networks}
\label{tab:multi}
\end{table*}

\noindent \textbf{Sampling baseline: } \label{sec:baselines} For selecting the best baseline, we tested four sampling based graph discovery methods which are used in \cite{wilder2018maximizing, valente2007identifying}: CHANGE, SNOWBALL, RECOMMEND and RANDOM-GREEDY. We found that Random greedy, Snowball and recommend are 3.9\%, 42.1\% and 41.7\% worse than CHANGE respectively. Hence, \textit{we only use CHANGE as the baseline.}
CHANGE is a recent state-of-art method that was used for effective HIV intervention campaign. It uses a simple yet powerful sampling method: \emph{For each of the random seeds, we query one of its neighbors picked at random.} The model is inspired by \emph{friendship paradox} which states that the expected degree of a random node's neighbor is larger than the expected degree of a random node. 


\noindent \textbf{Environment parameters:}
We assume that information flow is modelled by independent cascade model \cite{kempe2005influential}.
We assign it the same parameters to the model as done in \cite{wilder122018maximizing} since it is similar to the real-world deployment setting.
We fix the diffusion probability $p$ for edges as 0.1. 
We fix the budget for the number of nodes to be activated at 10.
Similar to setting in \cite{wilder122018maximizing}, before we start network discovery, we are given 5 random seed nodes $\mathcal{R}$ and their neighbourhood is revealed. We have $T=5$ queries to discover the graph $G_T$ using which we find the 10 nodes to activate. \\ 

\noindent \textbf{Performance metrics: } We employ the greedy algorithm, which we denote as the oracle $\mathcal{O}$, on the discovered graph to pick the nodes to activate. We remark that greedy gives the optimal approximation ratio unless $P=NP$ \cite{kempe2005influential}. 
$OPT$ is the influence score when we have the entire network for $\mathcal{O}$ to choose from. This value can be represented as $I_{G^*}(\mathcal{O}(G^*))$ (note that we approximate $OPT$ via the greedy algorithm since exact optimization is NP-hard.)
We use the following performance metrics to validate our models:

(1) \emph{influence score} or \emph{influence reward}: We deploy our model on graphs from the test set and consider the \emph{average number of nodes influenced} over 100 runs as the performance metric.
    
    (2) \emph{improve percent}: percentage reduction with respect to the gap between $OPT$ and a given method. (This is the scaled reward from Equation \ref{eqn:scale}). This 
is useful to aggregate performance over multiple networks since the range of values for actual \emph{influence scores} for each network vary depending on network properties.

\section{Results}
The policies learned through reinforcement learning by our agent results in a significant increase in the number of nodes influenced.
Table \ref{tab:best} shows the scores of the best-performing variant of our agent on each test set. Geometric-DQN improves the total number of nodes reached by 7-26\% compared to CHANGE, allowing us to substantially improve the effectiveness of influence maximization interventions without increasing the budget for network discovery. 

\begin{table}[!h]
\centering
\begin{tabular}{|l|c|}
\toprule
\textbf{Network Family}  & \textbf{improve \%} \\
\midrule
Rural                   & 23.76\\
Animals                  & 26.6\\
Retweet                 & 19.7\\
Homeless                & 7.91\\
\bottomrule
\end{tabular}
\caption{Summary of best scores averaged over test networks}
\label{tab:best}
\end{table}

In the following subsections, we discuss multiple ways we can improve robustness of training in the face of uncertainties such as choosing the right training graph or overcoming lack of actual real-world network to train on. First, we show that simultaneously training with multiple networks is on average better than using a single network. Then, we show the effectiveness of synthetic graphs both as an effective substitute for real networks and as a valuable method for data augmentation. \\
\noindent \textbf{Training on Individual network vs Multiple networks:}
We can directly use the training networks to discover efficient policies for sampling from the unknown network of similar domain.
One way to do this is to pick one of the train networks to train on and this indeed outperforms the baseline.
However, using scaled rewards (eqn: \ref{eqn:scale}) allows us to leverage information from multiple networks simultaneously. As shown in table \ref{tab:multi}, \textit{training with multiple networks gives better performance gains compared to average scores obtained with training on single networks on all but one.} \\
\noindent \textbf{Synthetic graphs can be valuable.} Synthetic graphs can be useful substitutes when we have limited access to real-world social networks.
Since synthetic graphs are cheaper alternatives to collecting more real-world data, they can also be used along with available networks for training.
We performed two different experiments to validate the usefulness of synthetic graphs: 1) Use only the synthetic graphs generated for training, 2) Use synthetic graphs along with training dataset. 
For each training graph we generated 5 synthetic graphs from SBM model ( as discussed in Section \ref{sec:syn}).
During training, we use synthetic graphs with probability 0.5 for an episode. Otherwise, we sample from the training dataset. We empirically found that this method gave best performance.

\begin{figure}[h]
    \centering
    \begin{subfigure}{.25\textwidth}
      \centering
      \includegraphics[width=\linewidth]{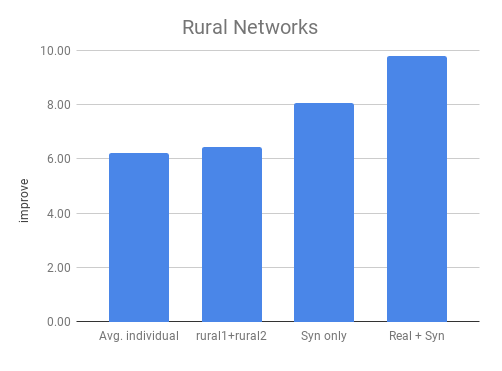}
    \end{subfigure}%
    \begin{subfigure}{.25\textwidth}
      \centering
      \includegraphics[width=\linewidth]{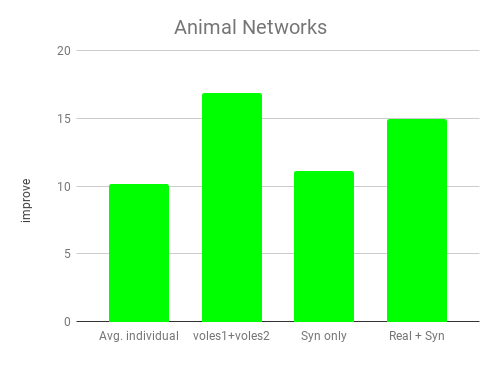}
    \end{subfigure}
    \begin{subfigure}{.25\textwidth}
      \centering
      \includegraphics[width=\linewidth]{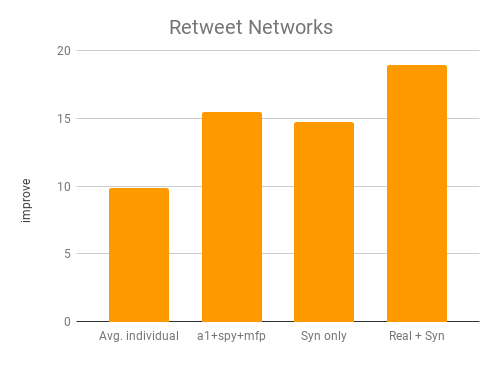}
    \end{subfigure}%
    \begin{subfigure}{.25\textwidth}
      \centering
      \includegraphics[width=\linewidth]{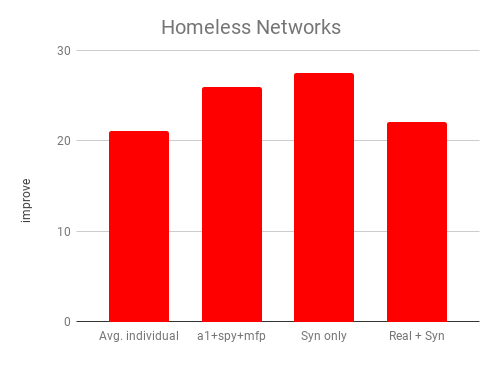}
    \end{subfigure}
   \caption{Variation of average \emph{improve percent} using synthetic data. Syn only: Only synthetic graphs for training. Real+Syn: Use both synthetic and actual train graphs}
   \label{fig:syn}
\end{figure}

During training, to generate a synthetic graph we sampled a graph from training set, fit a SBM (or SSM for the retweet dataset) model, and sampled a synthetic graph.
We compare \emph{improve percent} for different families of graphs using these two settings along with other settings in figure \ref{fig:syn}.
We see that the \emph{improve percent} scores for models using only synthetic graphs are better than the average of the scores of models trained from individual networks. \textit{This indicated that training on synthetic graph alone is a more robust method than training on individual real networks.}
For Homeless, Retweet and Rural networks, data augmentation improved the performance on the model trained on only training graphs.
In the case of Homeless networks using only synthetic graphs give the highest score on average. This may indicate that the SBM model generates graphs that model some of the Homeless networks very well. 
These experiments suggest that using synthetic graphs when we don't have access to real graph datasets can be very effective.

\section{Insights on policy learnt}
\label{sec:policy}
We observe some of the characteristics of the policy learnt by the Geometric-DQN models and investigate some interpretable quirks of the policies that lead to improvement over CHANGE.

\noindent \textbf{Size of the discovered graph} In general, we note that the Geometric-DQN policies almost always discover larger graphs than CHANGE though there is no correlation between influence scores and size of discovered graphs by different Geometric-DQN policies. Therefore, we further explore the properties of the nodes selected by the policy.

\noindent \textbf{Observations on node selection}
We observed two frequent events, labelled \emph{O1} and \emph{O2}, during deployment on test networks:\\
\emph{O1}: The next node to be selected has \emph{minimum} degree\\
\emph{O2}: The node selected is from set of nodes most recently discovered.\\
We found that almost all the time, if \emph{O2} occurs, \emph{O1} also does. 

\begin{figure}[h]
    \centering
    \includegraphics[width=0.3\textwidth]{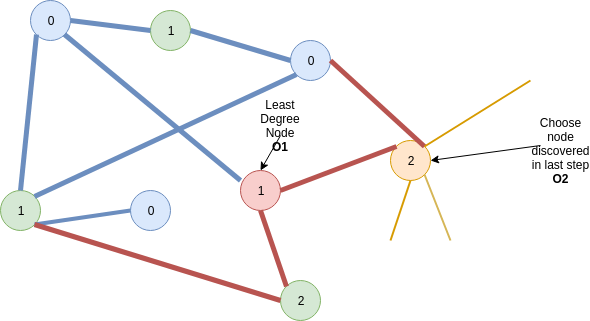}
    \caption{A toy example demonstrating observations \emph{O1} and \emph{O2}. The number on top of nodes denote the time-step at which they where discovered. Red nodes  satisfy O1. Orange nodes satisfy O2. Green nodes are other potential actions. Blue denotes seed nodes}
    \label{fig:steps}
\end{figure}

\begin{table}[h]
\centering
\scalebox{1}{
\begin{tabular}{|l|c|c|c|c|}
\toprule
\textbf{Graph} & \textbf{CHANGE} & \textbf{H1}  & \textbf{H2} & \textbf{Geometric-DQN} \\
\midrule

rural3         & 17.4              & 17.36      & 17.3          & \textbf{18.95}        \\
rural4         & 31.5              & 32.39      & 32.6          & \textbf{35.7}         \\\hline
voles3          & 33.7             & 38.7       & 39.6          & \textbf{45.8}        \\
voles4          & 58.9              & 61.9      & 72.8          & \textbf{80.2}         \\\hline
damascus        & 95.2             & 93.7       & 104.9         & \textbf{119.3}         \\
israel          & 30.6                 & 30.37  & 34.2          & \textbf{43.6}        \\\hline
b1              & 19.1              & 19.0      & 19.4          & \textbf{20.2}         \\
spy2            & 16.01             & 15.877    & 16.4          & \textbf{17.4}        \\
\bottomrule
\end{tabular}
}
\caption{Comparisons of scores of heuristics with baselines and best of Geometric-DQN models for each graph}
\label{tab:heu}
\end{table}

\paragraph{Heuristics based on observations}
To verify that the behaviours $O1$ and \emph{O2} were beneficial for our task, we devised two heuristics.\\
    H1: \emph{Query only from the nodes with \emph{minimum} degree in sub-graph}\\
    H2: \emph{Query only from the nodes with minimum degree from the set of node discovered in the \emph{previous} step. If no new nodes are discovered in the previous step, choose unqueried node in the discovered graph.}\\
We break all ties by choosing uniformly at random. The performance of the heuristics is summarized in Table \ref{tab:heu}.
We observe that H2 outperforms CHANGE by huge margins for Animals, Homeless and Retweet networks, whereas H1 performs comparable to CHANGE in all networks.
Geometric-DQN models still perform much better than heuristics. This indicates that the policies learned by Geometric-DQN that are adapted to the specific class of training networks are more nuanced than the heuristics we designed.

\noindent \textbf{Properties of selected nodes}
To further investigate why the heuristics and Geometric-DQN policy performs better, we look at \textit{degree centrality} and \textit{betweenness centrality} of the nodes queried in the true underlying graph.
We call betweenness and degree centrality of a node computed on the true graph as its \emph{true betweenness centrality} and \emph{true degree centrality} respectively.
 Picking nodes with high true degree centrality allows access to a larger number of nodes during discovery.
 For network discovery, nodes of high true betweenness centrality could act as a bridge between different strongly connected communities of nodes for further exploration. In relation to influence maximization, nodes with true high betweenness centrality can allow the flow of information between parts of the network which would otherwise be hard to access.

\begin{figure}[h]
    \centering
    \begin{subfigure}{.25\textwidth}
      \centering
      \includegraphics[width=\linewidth]{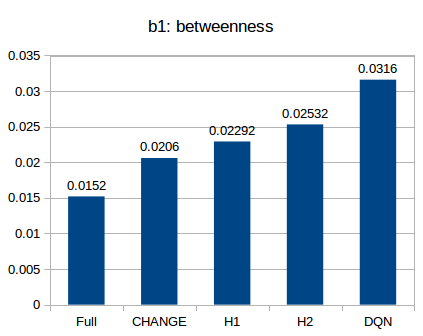}
    \end{subfigure}%
    \begin{subfigure}{.25\textwidth}
      \centering
      \includegraphics[width=\linewidth]{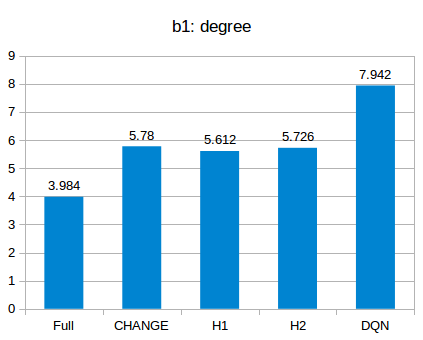}
    \end{subfigure}
    \begin{subfigure}{.25\textwidth}
      \centering
      \includegraphics[width=\linewidth]{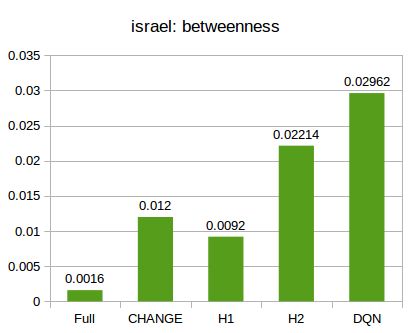}
    \end{subfigure}%
    \begin{subfigure}{.25\textwidth}
      \centering
      \includegraphics[width=\linewidth]{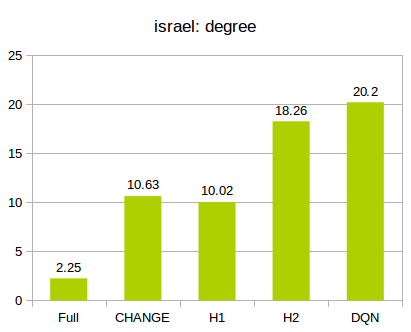}
    \end{subfigure}
    \caption{Average true betweenness (R) and degree(L) centrality of Full graph, nodes queried by CHANGE, H1, H2 and best Geometric-DQN  }
    \label{fig:deg}
\end{figure}

In particular, we report studies on \emph{b1}, an interaction network depicting real-world community structure and \emph{israel}, a retweet network with sparsely interconnected star-graphs.
We compare the true degree centrality and true betweenness centrality of queried nodes using CHANGE, Geometric-DQN and the heuristics discussed above.
As we can see from Figure \ref{fig:deg}, compared to other models, the Geometric-DQN can recognize nodes with high true degree centrality and true betweenness centrality. H2 also picks nodes with higher true betweenness centrality than CHANGE, but does not improve in terms of average true degree centrality.

\begin{figure}[h]
    \centering
    \begin{subfigure}{.25\textwidth}
      \centering
      \includegraphics[width=\linewidth]{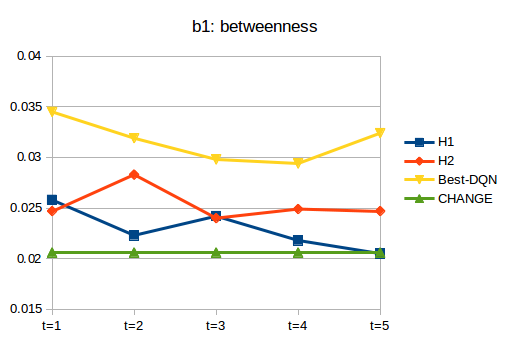}
    \end{subfigure}%
    \begin{subfigure}{.25\textwidth}
      \centering
      \includegraphics[width=\linewidth]{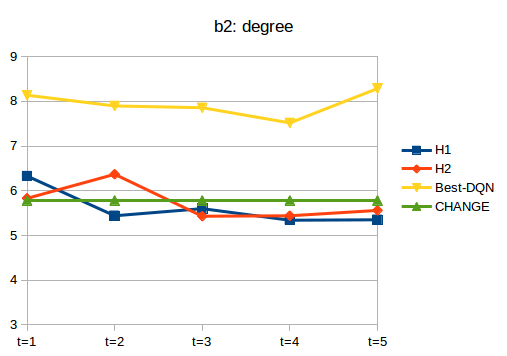}
    \end{subfigure}
    \begin{subfigure}{.25\textwidth}
      \centering
      \includegraphics[width=\linewidth]{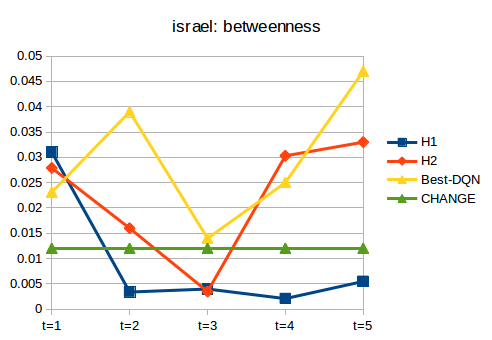}
    \end{subfigure}%
    \begin{subfigure}{.25\textwidth}
      \centering
      \includegraphics[width=\linewidth]{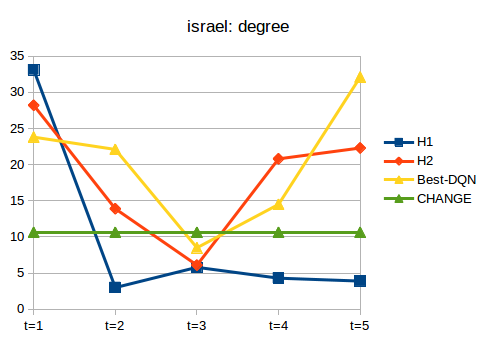}
    \end{subfigure}
   \caption{Average true betweenness(L) and degree(R) centrality across timesteps for H1,H2 and best Geometric-DQN. (We have added corresponding CHANGE values for reference)  }
   \label{fig:time}
\end{figure}
We further investigate these properties vary across timesteps for H1, H2 and best Geometric-DQN model (see Figure \ref{fig:time}).
On average, Geometric-DQN finds nodes of high full betweenness centrality and degree centrality, especially in the last query.
This may indicate that Geometric-DQN has leveraged the Deepwalk embeddings as well as the learned graph embeddings to find complex higher-order patterns in the graphs that enable it to find such nodes.

\section{Related works}
 \paragraph{Influence Maximization and network discovery} 
The existing literature on algorithmic aspects of influence maximization problems \cite{kempe2003maximizing,jung2012irie,chen2010scalable,tang2014influence,li2018influence} build on a greedy strategy for influence maximization, which iteratively adds the node with the largest marginal contribution to the objective until the budget is reached. 
Our work is situated along a more recent, largely orthogonal axis: developing algorithms which reduce the data collection requirements needed to deploy influence maximization in the real world. Wilder et al.\ \cite{wilder2018maximizing} introduced the exploratory influence maximization problem, where the goal is to sample nodes from the network such that an influential seed set can be selected. They propose an algorithm motivated by community structure and prove theoretical guarantees for graphs from the stochastic block model. \ \cite{wilder2018end} introduced CHANGE, a more practical algorithm based on the friendship paradox \cite{feld1991your}.


\paragraph{Reinforcement Learning for Combinatorial Problems on graphs}
There has been some recent work on the idea of using deep reinforcement learning methods to automatically learn heuristics to solve combinatorial problems on graphs that generalize well to graphs from a similar distribution.
E.g., \cite{Bello2016NeuralCO} use policy gradient methods to solve the Travelling Salesman Problem (TSP) and  \cite{Khalil2017LearningCO} use Q-learning to solve Minimum Vertex Cover, Maxcut and TSP. 
In contrast to these works, our setting deals with unknown, partially observed graphs, and we train an \emph{adaptive} policy for deciding where in the graph to query based on what as been observed so far. This distinction in problem setting renders earlier architectures ineffective for our problem because the state and action representations change as more information about the graph is uncovered.


\section{Discussion and conclusion}
We introduce a novel deep Q-learning based method to use structural properties of the available social networks to learn effective policies for the network discovery problem to maximize influence on undiscovered social networks.
Our method exploits structural properties of the networks and use synthetic graphs as a robust alternative to training on real-world graphs or can be used along with training graphs as data augmentation technique. We showed that our trained models outperform current state-of-the-art algorithms on social networks across different domains and showed 7-23\% improvement over CHANGE (a SOTA sampling algorithm). 
Our method could be extended to other graph discovery problems by altering the reward function.

The graph embeddings learned identified pick nodes with high betweenness centrality with respect to the entire network, which was key to discovering important portions of the network.
A detailed investigation into complex patterns learned by our model could reveal insights about structural properties of social network pertaining to the identification of influential nodes.

\clearpage
\bibliographystyle{ACM-Reference-Format}  
\bibliography{references}  

\end{document}